\documentclass[11pt,twoside]{article}
\usepackage{asp2010}

\resetcounters

\begin{document}

\title{High-resolution spectropolarimetric observations of hot subdwarfs}
\author{P.~Petit$^1$, V.~Van Grootel$^2$, S.~Bagnulo$^3$, S.~Charpinet$^1$, G.A.~Wade$^4$, E.M.~Green$^5$
\affil{$^1$Institut de Recherche en Astrophysique et Plan\'etologie, Toulouse, France}
\affil{$^2$Institut d'Astrophysique et de G\'eophysique, Li\`ege, Belgium}
\affil{$^3$Armagh Observatory, Armagh, Northern Ireland, UK}
\affil{$^4$Royal Military College of Canada, Kingston, Canada}
\affil{$^5$Steward Observatory, University of Arizona, USA}}

\begin{abstract}
We report on high-resolution spectropolarimetric observations of the hot subdwarf stars HD 76431 and Feige 66, using the ESPaDOnS echelle spectropolarimeter at CFHT. We compute cross-correlation Stokes $I$ and $V$ line profiles to enhance the signal-to-noise ratio. We then average all available cross-correlation profiles of each star to further decrease the noise level. Although both targets were previously reported to host kilo-gauss magnetic fields, we do not derive any evidence of large-scale photospheric fields from our sets of observations, in spite of tight error bars on the longitudinal field of the order of 60 gauss for HD 76431 and 200 gauss for Feige 66. A new analysis of FORS1 observations of HD 76431, which provided the basis for the original claim of field detection, confirms the absence of any detectable Zeeman signature, with an error bar of about 100 gauss on the longitudinal magnetic field. 
\end{abstract}

\section{Selected targets}

The two stars selected for the present study are previously identified subdwarf stars for which magnetic field detections were reported. HD~76431 has an effective temperature $T_{\rm eff} = 31,000$~K and a logarithmic surface gravity $\log~g = 4.5$ (Ramspeck et al. 2001). Although this object is listed as an sdB star in several studies, the relatively low surface gravity (confirmed by several authors) is rather indicative of a post-asymptotic branch star.  O'Toole et al. (2005) report for HD 76431 a magnetic field detection using the VLT/UT1 with the FORS1 spectropolarimeter, corresponding to a longitudinal field component of $-1096\pm 91$~G. Feige 66 is an sdO star with an effective temperature of 34,500~K and a logarithmic surface gravity $\log~g = 5.83$ (O'Toole et al. 2006), for which a field detection was reported by Elkin (1996), with a photospheric field strength varying between -1300~G and +1750~G (with error bars of the order of 200-500~G).

We revisit here these two targets, using the ESPaDOnS spectropolarimeter.   
 
\section{Spectropolarimetric data sets}

\begin{figure}
\centering
\includegraphics[width=9.5cm]{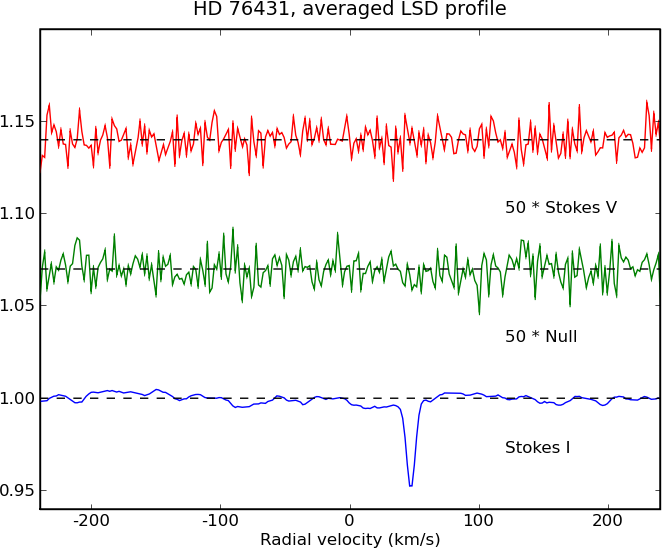}
\vspace{1cm}
\includegraphics[width=9.5cm]{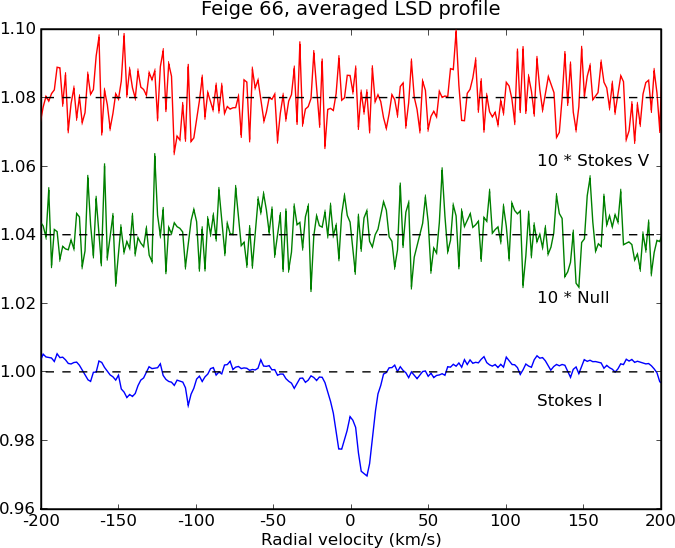}
\caption{Phase-averaged LSD line profiles derived from ESPaDOnS observations of HD 76431 (top panel) and Feige 66 (bottom panel). The lower (blue) curve is the intensity (Stokes $I$) profile. The top (red) line is the circularly-polarized (Stokes $V$) line profile. The middle (green) curve is a control (null) profile that should not contain any signature. For display clarity, the Stokes $V$ and null profiles are shifted vertically and multiplied by a factor of 50 (HD 76431) and 10 (Feige 66).}
\label{fig:espadons}
\end{figure}

Repeated observations of HD 76431 and Feige 66 were carried out with the ESPaDOnS spectropolarimeter at Canada-France-Hawaii Telescope. With this instrument, we benefit from the coverage, in a single exposure, of the whole spectral domain from 370~nm to 1,000~nm, at a spectral resolution of about 65,000  (Petit et al. 2003). Using ESPaDOnS in polarimetric mode, we simultaneously record spectra in intensity (Stokes $I$) and circular polarization (Stokes $V$).

This instrumental setup was employed to gather 17 observations of HD 76431, from 2009 Oct. 02 to 2010 Feb. 02, adopting an integration time of 2,400~s. Feige 66 was observed 13 times between 2010 Jan. 25 and 2010 Jun. 02, with an exposure time of 3260~s. 

\section{Magnetic field measurements using ESPaDOnS}

Using two lists of spectral lines matching stellar photospheric models for the spectral types of HD 76431 and Feige 66 (91 spectral lines for HD 76431 and 39 spectral lines for Feige 66), we calculate from every spectrum an average photospheric line profile, using the Least-Square-Deconvolution multi-line technique (hereafter LSD, Donati et al. 1997). Thanks to this cross-correlation method, the noise level of the mean Stokes $V$ line profiles  is reduced by a factor of 2.5 to 6 (for Feige 66 and HD 76431, respectively) with respect to the initial spectrum. The resulting noise level of the mean profiles is in the range $5\times10^{-4}~I_{\rm c}$  to $2\times10^{-3}~I_{\rm c}$ (where $I_{\rm c}$ denotes the intensity of continuum) for HD~76431, $2\times10^{-3}~I_{\rm c}$ to $1.5\times10^{-2}~I_{\rm c}$ for Feige 66. The analysis of individual LSD profiles does not lead to any detection of Zeeman signatures at specific rotational phases, with upper limits of 100~G and 500~G (for HD 76431 and Feige 66, respectively, using for each star the LSD profile with the highest available S/N).  

To further decrease the noise, we calculate a temporal average of all available observations of each star (following, e.g.,  Petit et al. 2011), thereby reaching a S/N of 7,000 and 1,400 for HD 76431 and Feige 66, respectively (Fig. 1). We do not detect any Zeeman signature in the averaged profiles, with a false-alarm probability (inside the line profile) of 0.2 for HD 76431 and 0.3 for Feige 66. The resulting phase-averaged longitudinal field component is equal to $12\pm 55$ G and $240\pm 200$ G, for HD 76431 and Feige 66, respectively. Note that the double-peaked Stokes I profile obtained for Feige~66 suggests that this star is a SB1.

\section{Revisiting FORS~1 observations of HD~76431}

\begin{figure}
\centering
\includegraphics[width=12cm]{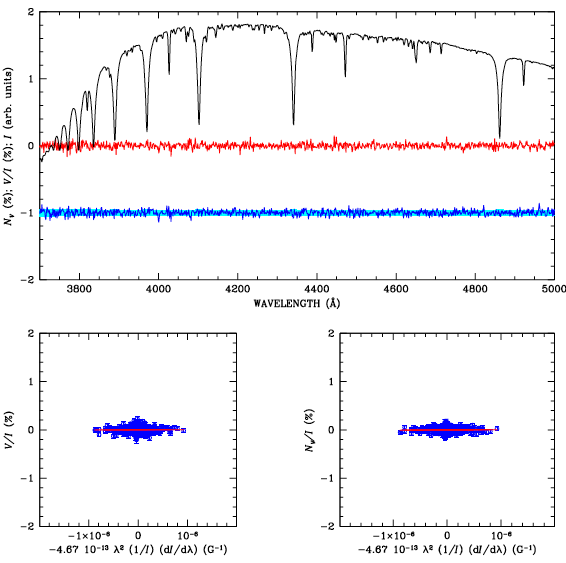}
\caption{FORS spectrum of HD 76431. The top panel shows Stokes $I$ (black line), V/I (red solid line centred about 0), and the null profile (blue solid line, offset by -1\% for display purpose). The slope of the interpolating lines in the bottom panels gives the mean longitudinal field from $V/I$ (left panel) and from the null profile (right panel, the latter being expected zero).}
\label{fig:fors}
\end{figure}

FORS1 observations of HD 76431 were obtained on 2004-02-23 at UT 06:06 with grism 600B. The slit width was set to 0.5" for a spectral resolution of about 1,400. The exposure time was 1,800s, and the peak signal-to-noise ratio was 2,675 per \AA. From this data set, O'Toole et al. (2005) reported the detection of a surface magnetic field, with a line-of-sight component equal to $-1096\pm91$ gauss. We re-reduced the observations along the lines explained by Bagnulo et al. (2002, 2006), namely by estimating the mean longitudinal magnetic field with the relationship~:

\begin{equation}
V/I = -4.67 \times 10^{-13}\lambda^2\frac{1}{I}\frac{{\rm d}I}{\rm d \lambda}B_l
\end{equation}

\noindent We find that the mean longitudinal field from H Balmer lines is $-54\pm110$ G. From the He and metal lines, we measure $252\pm117$ G, and from the full spectrum we measure $44\pm88$ G, in agreement with the absence of detection obtained from ESPaDOnS spectra. The magnetic field measured from the null profiles was always found to be consistent with zero. The top panel of Figure 2 shows Stokes $I$ (black solid line, in arbitrary units, and not corrected for the instrument response), $V/I$ (red solid line centred about 0), and the null profile (blue solid line, offset by -1\% for display purpose). Light blue bars represent the error bars of $V/I$. The slope of the red interpolating lines in the bottom panels gives the mean longitudinal field from $V/I$ (left panel) and from the null profile (right panel, the latter being expected zero). A thorough review of the FORS1 data archive is presented by Bagnulo et al. (2011, in press). 

\section{Discussion}

Repeated series of observations obtained with the ESPaDOnS high-resolution spectropolarimeter, analysed with a cross-correlation technique aimed at optimizing the extraction of Zeeman signatures, we place tight upper limits of  60 G and 200 G on the surface magnetic fields of HD 76431 and Feige 66, respectively. 

These findings are at odds with the study of Elkin (1996), who reported the discovery of a kilo-gauss magnetic field at the surface of Feige 66. Our result is also discrepant with the work of O'Toole et al. (2005) who also report the presence of a kG field on HD 76431 and other hot subdwarfs. A new analysis of the data set collected by O'Toole does not confirm their early claim for a field detection, suggesting that the previously reported Zeeman signatures were artifacts generated by the data reduction pipeline. We suggest that the same may apply for Feige 66.

The present study calls into question the admitted occurrence of strong magnetism in hot subdwarfs. New spectropolarimetric  observations of a larger sample of sdO and sdB targets is required to determine the fraction of such objects hosting strong surface fields.     

\nocite{*}

\bibliography{author}

\end{document}